\documentclass[journal=jpccck,english,manuscript=article]{achemso}
\usepackage{subfigure}
\usepackage{babel}

\author{\textbf{O. Godsi}}
\affiliation{\textbf{Schulich Faculty of Chemistry, Technion - Israel Institute of Technology, Technion City, Haifa 32000, Israel.}}

\author{\textbf{G. Corem}}
\affiliation{\textbf{Schulich Faculty of Chemistry, Technion - Israel Institute of Technology, Technion City, Haifa 32000, Israel.}}
\author{\textbf{T. Kravchuk}}
\affiliation{\textbf{Schulich Faculty of Chemistry, Technion - Israel Institute of Technology, Technion City, Haifa 32000, Israel.}}
\author{\textbf{C. Bertram}}
\affiliation{\textbf{Ruhr-Universit\"at Bochum, Lehrstuhl f\"ur physikalische Chemie I, D-44780 Bochum, Germany.}}
\author{\textbf{K. Morgenstern}}
\affiliation{\textbf{Ruhr-Universit\"at Bochum, Lehrstuhl f\"ur physikalische Chemie I, D-44780 Bochum, Germany.}}
\author{\textbf{H. Hedgeland}}
\altaffiliation{London Centre for Nanotechnology, University College London, London WC1H 0AH, United Kingdom.}
\affiliation{\textbf{The Cavendish Laboratory, J. J. Thomson Avenue, Cambridge, CB3 0HE, United Kingdom.}}
\author{\textbf{A. P. Jardine}}
\affiliation{\textbf{The Cavendish Laboratory, J. J. Thomson Avenue, Cambridge, CB3 0HE, United Kingdom.}}
\author{\textbf{W. Allison}}
\affiliation{\textbf{The Cavendish Laboratory, J. J. Thomson Avenue, Cambridge, CB3 0HE, United Kingdom.}}
\author{\textbf{J. Ellis}}
\affiliation{\textbf{The Cavendish Laboratory, J. J. Thomson Avenue, Cambridge, CB3 0HE, United Kingdom.}}
\author{\textbf{G. Alexandrowicz}}
\email{ga232@tx.technion.ac.il}
\affiliation{\textbf{Schulich Faculty of Chemistry, Technion - Israel Institute of Technology, Technion City, Haifa 32000, Israel.}}

\title{How Atomic Steps Modify Diffusion and Inter-adsorbate Forces: Empirical Evidence From Hopping Dynamics in Na/Cu(115).}

\begin{document}

\begin{abstract}

We followed the collective atomic-scale motion of Na atoms on a vicinal Cu(115) surface within a time scale of pico to nano-seconds using helium spin echo spectroscopy. The well defined stepped structure of Cu(115) allows us to study the effect that atomic steps have on the adsorption properties, the rate for motion parallel and perpendicular to the step edge and the interaction between the Na atoms. With the support of a molecular dynamics simulation we show that the Na atoms perform strongly anisotropic one dimensional hopping motion parallel to the step edges. Furthermore, we observe that the spatial and temporal correlations between the Na atoms which lead to collective motion are also anisotropic, suggesting the steps efficiently screen the lateral interaction between Na atoms residing on different terraces.  
\end{abstract}

Single atomic steps are the dominant surface defect, even on a carefully
prepared surface. It is widely excepted that atomic steps play an important role in surface diffusion of add-atoms and adsorbates and hence their existence and density is expected to modify the growth mode of thin-layers, heterogeneous
catalysis and many other surface systems which are limited by the diffusion rate. Furthermore, it has
been hypothesized that atomic steps are responsible
for controversies and discrepancies between diffusion measurements performed with techniques that have different
spatial resolutions\cite{Antczak_book,ala-nissila_collective_2002}. 

Significant efforts have been made to measure the effect steps have on surface diffusion of adsorbates over the years. Macroscopic techniques which measure variations in coverage on a micron scale have shown the cumulative effect of a large number of atomic steps, clearly demonstrating the anisotropic nature of the motion and the changes in the activation energies for diffusion \cite{Ma1999L661,Ma1999} .Unfortunately, experimental data for the atomic-scale motion of an adsorbate near a step edge is particularly scarce. The tendency of adsorbates to be trapped at step sites makes it difficult, if not impossible, to study the diffusion near a step using low temperature scanning tunneling microscopy due to its limited dynamical range.  Under certain conditions, adsorbates can be released from the steps using energetic photons in pump-probe experiments, providing valuable insight into the dynamics of the excited adsorbate\cite{Stepan2005, backus_real-time_2005}. In contrast, when it comes to studying how the steps modify the individual and collective atomic-scale motion of adsorbates in thermal equilibrium, the experimental data base is essentially non-existent leaving us with an understanding based solely on theoretical  approaches. 
 \cite{ala-nissila_collective_2002,chvoj2006}  
  
In this work, we present measurements of the atomic scale motion of Na
atoms on a Cu(115) surface. The well defined geometry of a densely stepped vicinal surface such as Cu(115) \cite{diehl2015}, provides a controlled environment
for studying the role atomic steps have on diffusion. Since the motion of Na on Cu(100)
surface has been thoroughly studied in the last two decades \cite{ellis_observation_1993, graham_determination_1997,cucchetti_diffusion_1999,alexandrowicz_onset_2006},
working with the Cu(115) surface provides an excellent opportunity
to focus on the changes induced by the presence of atomic steps. The transition from the relatively flat Cu(100) surface to a stepped surface raises several questions and unknowns, in particular: which are the available adsorption sites on this surface and to what extent do the atomic steps modify the diffusion, both parallel and perpendicular to the step direction. Another question, which is typically inaccessible to experiments, is whether the steps affect the lateral interactions between the adsorbed atoms i.e. the collective diffusion process. We combine experimental data with molecular dynamics (MD) simulations to address these questions and show that atomic steps radically change the nature of the atomic-scale diffusion.

The Cu(115) sample studied in this work is a vicinal surface with (100) terraces and
a very small step separation of 6.63$\dot{A}$. The schematic drawing in figure 1 illustrates
the geometry of the surface viewed from above and from the side. The
clean Cu(115) surface has been shown to be a stable vicinal surface
below the roughening transition (380K)\cite{fabre1987}.
Adsorption of alkali
metals at high coverages and temperatures can initiate a surface
reconstruction\cite{Braun1995265}. In this study, we have limited
ourselves to relatively low coverages and temperatures where the surface
does not undergo any major reconstruction, verified using helium diffraction
before and after Na adsorption. 
\begin{figure}
    \includegraphics[width=60mm]{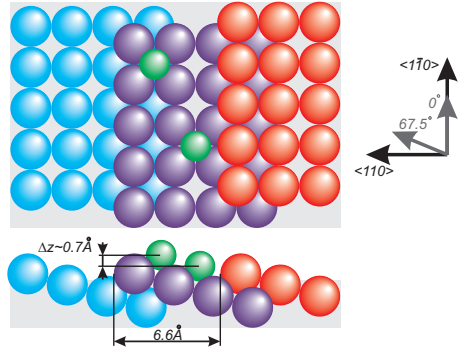}
    


\caption{\label{cap:Cu115schematic}Schematic drawing of the Cu(115) surface
viewed from above and from the side, and a definition
for the $0^{\circ}$ and $67.5^{\circ}$ azimuths referred to in the
text. The copper atoms of the different layers are drawn with different
colored spheres, smaller green spheres illustrate a hypothetical scenario of
Na atom moving in between hollow cites, emphasizing that such a motion would also have a component perpendicular to the surface.}
\label{figure1}
\end{figure}

The experimental technique used in this study is Helium Spin Echo
(HeSE) spectroscopy. In a HeSE experiment the nuclear spin of $^{3}$He
is manipulated using magnetic fields resulting in an atom interferometer
setup, capable of measuring atomic-scale motion in reciprocal space on a unique
time range of pico to nano-seconds. In the text below, we provide
a brief description of the quantities measured by this technique; a detailed explanation can be found elsewhere \cite{alexandrowicz_helium_2007,jardine_studying_2009}.

Within the kinematic scattering approximation, the HeSE apparatus measures the
intermediate scattering function, ISF,  $I\left(\Delta\vec{k},t\right)\propto<\sum\limits_{i}\sum\limits_{j} e^{-i\Delta\vec{k}\cdot\vec{R}_{i}\left(0\right)}e^{-i\Delta\vec{k}\cdot\vec{R}_{j}\left(t\right)}>$.
$\Delta\vec{k}$ is the momentum exchanged during the scattering of
the helium atom (divided by $\hbar$) consisting
of a component parallel to the surface,
$\bar{\Delta K}$, and a component perpendicular to the surface, $\Delta k_{z}$
. $\vec{R}_{i}\left(0\right)$ and $\vec{R}_{j}\left(t\right)$ are
the position vectors of surface atoms (scattering centers) $i$ and
$j$ at times 0 and $t$ respectively, the double sum is over
all possible pairs of surface atoms and the brackets denote an
ensemble average\cite{ala-nissila_collective_2002}. When the coverages are low and
the interaction between the adsorbed species is negligible, the ISF
is dominated by the $i=j$ terms (self correlation). In this case, the diffusion of an adsorbed atom, $i$,
increases the difference between $\vec{R}_{i}\left(0\right)$ and
$\vec{R}_{i}\left(t\right)$ resulting in a random phase of the complex exponent (loss of spatial and temporal self-correlation) 
and a decay of the ISF to zero with increasing
time. When the interaction between the adsorbed atoms is non-negligible, the ISF
decay reflects the loss of self-correlation as well as the loss of correlation between pairs of adsorbates due to the dynamics. Hence, measuring correlation functions such as the ISF, does not restrict the data interpretation to a single parameter such as the tracer (self) or chemical (collective) diffusion coefficients, rather it supplies a full picture which allows to study both of these processes\cite{ala-nissila_collective_2002}. In many cases, including this work, the time dependence of the ISF is well approximated by
a simple exponential decay, $I\left(\Delta\vec{k},t\right)=Ae^{-\alpha t}$
, such that the $\Delta\vec{k}$ , temperature
and adsorbate coverage dependencies of the decay rates, $\alpha$,
supply a detailed picture of the microscopic diffusion process, the energy barriers
for diffusion, the frictional coupling to the surface and a unique opportunity to measure
inter-adsorbate forces\cite{alexandrowicz_onset_2006,jardine_studying_2009,hedgeland_measurement_2009}.

\begin{figure}
  \includegraphics[width=\columnwidth]{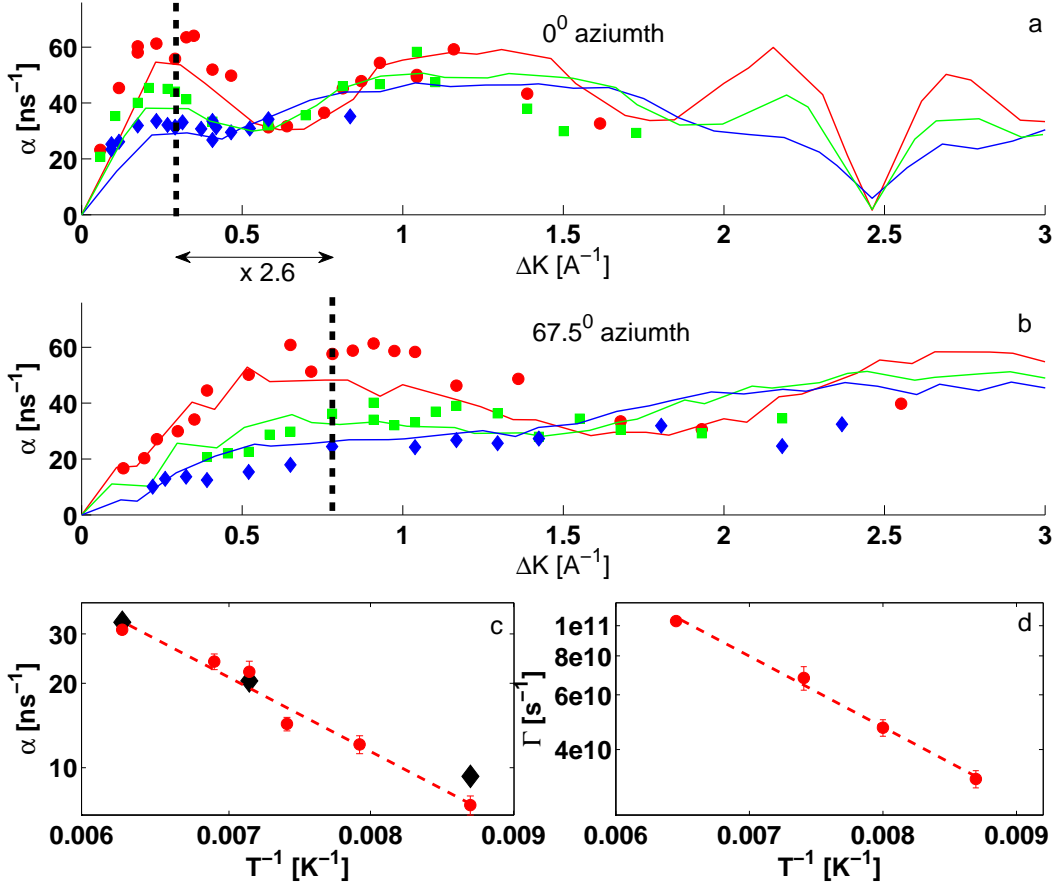}

\protect\caption{\label{cap:Data=000026fits}The full symbols show the
extracted decay rates, $\alpha$, of the measured ISF as function
of $\Delta K$, measured along the 0$^{\circ}$ azimuth (2a)
 and along the 67.5$^{\circ}$ azimuth (2b). Red circles, green squares and blue diamond symbols mark Na coverages of 0.095, 0.075 and 0.05 ML respectively. All measurements were
performed at 155K. The dashed vertical lines mark the approximate position of the correlation peak which is scaled by a factor of ~2.6 when moving from the 0$^\circ$ to the 67.5$^\circ$ azimuth.
The full lines show the MD simulation
results when using a high step-crossing barrier and an anisotropic
Na-Na interaction for the different coverages. Figure 2c  shows the temperature dependence of $\alpha$, measured along the 0$^{\circ}$ azimuth, $\Delta K=0.65$\AA$^{-1}$ at a Na coverage of 0.09ML. Red circular markers are the experimental values, the dashed red line is an Arrhenius fit with an activation energy of $53\pm5meV$, and the black diamond markers are results of the MD simulation using the best fit parameters described in the text.  Figure 2d shows the temperature dependent Na hopping rates extracted from the simulated trajectories.}
\end{figure}

In order to study the directionality of the surface diffusion we measured the ISF along two different crystal azimuths. A natural choice would be to measure parallel and perpendicular to the step edges, along the <$\bar{110}$> and <${110}$> crystal azimuths. On the flat surface these azimuths should give identical results, and any differences between the two could be used to quantify the effects steps have on the motion. However, along the <${110}$> direction the signal is overwhelmed by elastic scattering from the Bragg peaks of the stepped structure, hence, we chose a second azimuth which is 67.5$^\circ$ from <$\bar{110}$>, which is sufficiently far from the Bragg peaks yet still  provides insight into the anisotropic nature of the surface. The two azimuths are illustrated in figure 1 and are referred to in the text as the 0$^\circ$ and 67.5$^\circ$ azimuths.
The full symbols in figure 2a show the extracted decay rates, $\alpha$,
of the measured ISF as function of $\Delta K$, measured along
the 0$^\circ$ crystal azimuth. The measurements
were performed at 155K for Na coverages of 
0.05, 0.075 and 0.095 ML \footnote{ 1ML is defined as 1 Na atom per surface area
of 2.55\AA x2.55\AA for easier comparison with the Cu(100) surface}
. A few observations can be readily made from the measured decay rates.
(I) The decay rate at low $\Delta K$ values increases significantly
with coverage producing a peak followed by a dip: we will denote these features the correlation peak and the de Gennes narrowing dip\footnote{The de Gennes narrowing is termed in analogy to the features seen in neutron scattering from correlated liquids\cite{DeGennes1959825}, simply speaking the idea is that the repulsive forces lead to preferred length scales where the system is stable and correspondingly the energy width (or decay rate) is at a minimum at the corresponding wave vector values. A more detailed explanation of the phenomena is given in a review of the HeSE technique  \cite{jardine_studying_2009}}. The de Gennes narrowing dip is located at
approx, 0.65\AA$^{-1}$ for the highest coverage and both shifts
to lower $\Delta K$ values and becomes less pronounced as the coverage
is decreased. A similar coverage dependency was seen on Na/Cu(100) and is characteristic
of correlated motion, driven by long range repulsive dipole-dipole
interaction between the Na atoms (alkali atoms, like many other atomic and molecular adsorbates, transfer charge to the underlying metalic substrate resulting in significant dipole moments \cite{diehl1996}). (II) A second maximum can
be seen at approximately 1.2\AA$^{-1}$, a peak which was also seen
for Na/Cu(100). This peak matches the expected maximum position of
$\alpha$ in the Chudley Elliot model \cite{Chudley-Elliott}, if the Na atoms are
jumping between adjacent adsorption sites which follow the copper
atoms spacing of 2.55\AA. This observation is consistent with
the fact that the terraces on the Cu(115) have a (100) geometry. 
(III) For all three coverages the decay rates seem to reduce toward
zero as $\Delta K$ approaches zero. At this limit,
the only momentum exchanged upon scattering is perpendicular to the
surface. Correspondingly, the scalar product in the
ISF involves only motion perpendicular
to the surface. Thus, a vanishing $\alpha(\Delta K\rightarrow0)$ means
that the motion we measure is predominantly within the surface plane.

Observations (I ) and (II) tell us that Na atoms perform jump diffusion
with a typical jump distance equal to the lattice spacing and that
the Na atoms significantly repel each other. Observation (III) shows us that
the complex 3D motion which was seen at high coverages (>0.05ML) for Na/Cu(100)\cite{alexandrowicz_onset_2006}
plays a negligible role on this surface, within the coverage range
we measured. Furthermore, the third observation also provides insight
regarding the number of adsorption sites within the width of a single
terrace. In particular, one could hypothesize naively that the Na atoms  
occupy the hollow sites on both sides of the terrace (green spheres in figure 1) and jump between these sites, similarly to the motion of Na on a flat
Cu(100) surface. Due to the skewed nature of this surface, this type of motion should change both the lateral position and
the height of the atoms leading to a non-vanishing  $\alpha(\Delta K\rightarrow0)$, in  contrast to our experimental
observations. It is important to note that the helium atoms scatter from the electron density \cite{Benedek2010} and hence observation (III) on its own, does not completely rule out motion between adsorption sites on both sides of the terrace, if for some reason the electron density of this surface deviates substantially from the geometrical model shown schematically in figure 1.    


Further insight is obtained from measurements along the 67.5$^{\circ}$ azimuth. The full symbols
in figure 2b are the extracted decay rates, $\alpha(\Delta K)$, measured along the 67.5$^{\circ}$ azimuth at 155K for the same three coverages plotted
in figure 2a. It is immediately obvious that the $\Delta K$ dependence
is very different to that measured along the 0$^{\circ}$ azimuth,
i.e. the diffusive motion is clearly anisotropic. In fact, the curves in figure 2b approximately overlap those shown in figure 2a if one scales the two horizontal axes using a factor of 2.6; illustrated by the dashed vertical lines in figure 2a and 2b which mark the approximate position of the correlation peak for each of the two azimuths. A simple stretching of the $\Delta K$ axis between the two measurements
suggests 1d motion which is purely parallel
to the atomic steps. This can be understood from the fact that the dephasing of the ISF is given by the scalar product $\Delta\vec{k}\cdot\vec{R}$. Consequently, the projection of a 1d jump between adjacent adsorption sites onto the 67.5$^{\circ}$ direction is  $\cos(67.5)$~$=1/2.6$ shorter, stretching the  $\Delta K$ dependency by the reciprocal factor.

The simple scaling relation between the $\alpha(\Delta K)$ curves of the two azimuths, excludes significant motion perpendicular to the step edges which would produce additional Chudley-Elliot oscillations and a completely different $\alpha(\Delta K)$ shape\cite{Chudley-Elliott}. This finding strengthens our previous conclusion that the Na atoms are not jumping in between two different sites within
the width of the terrace, regardless of whether the surface corrugation probed by the helium atoms follows the schematic shown in figure 1 or not. Moreover, the energy barrier for jumping
between different terraces must be considerably higher than that for
motion in the parallel direction, resulting in a negligible contribution
to the decay rates, $\alpha$, from jumping in between adjacent terraces. 

\begin{figure}
 \includegraphics[width=\columnwidth]{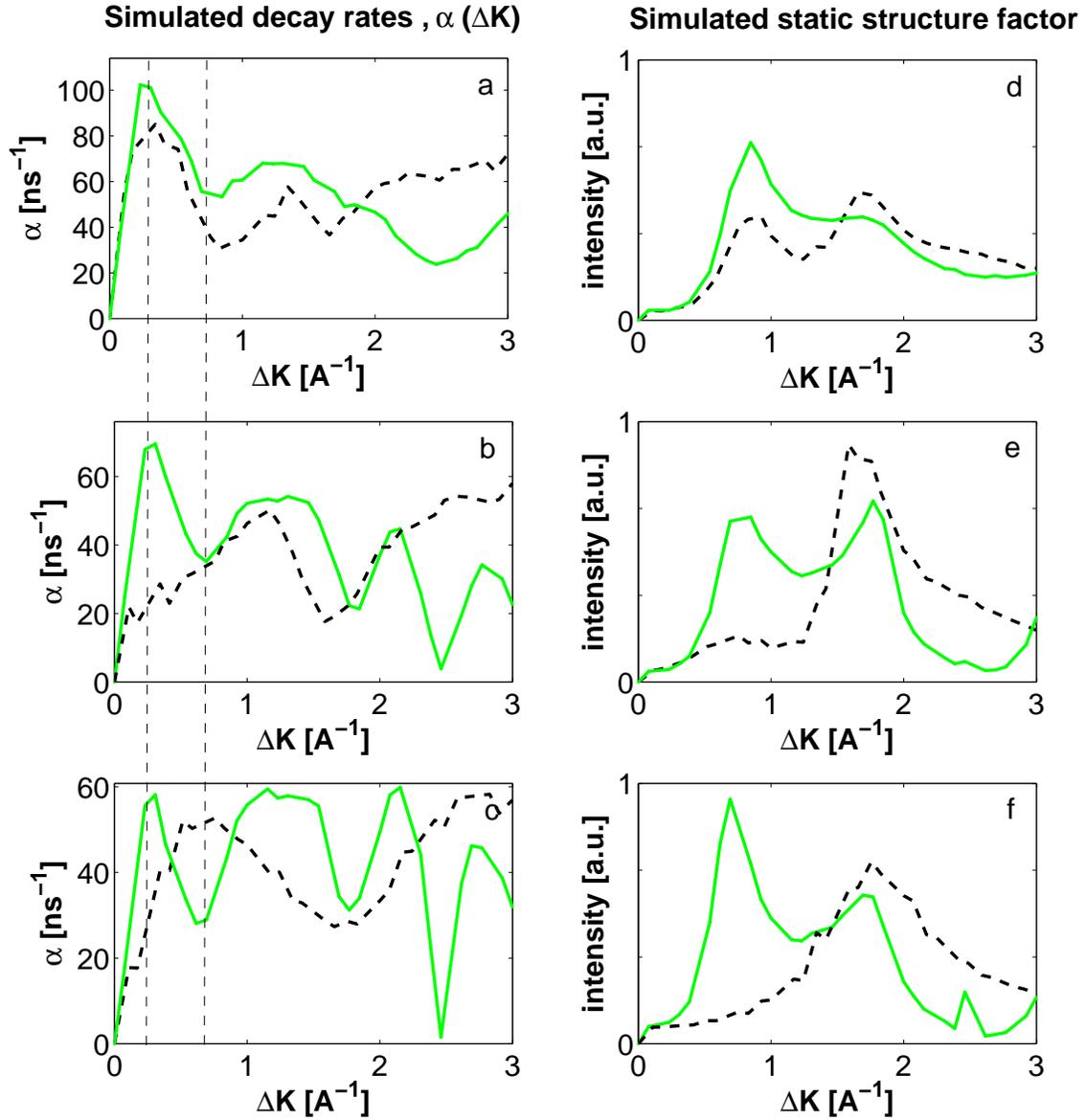}

\caption{\label{cap:Simulation_curves}MD simulations of $\alpha(\Delta K)$ curves
(left panels) and static structure factors (right panels), for a Na coverage of 0.095ML, T=155K, along the $0^{\circ}$ (solid green line) and $67.5^{\circ}$ (dashed black line)
azimuths respectively. Panels a \& d show results for a negligible (10meV)
step-crossing barrier, panels b \& e are for a high 
step-crossing barrier (70meV higher than the energy barrier for diffusion along the step edges) and panels c \& f are for the same
(70meV) step-crossing barrier when using an anisotropic interaction
model which screens the interaction between Na atoms on different terraces.}
\end{figure}

To complement these qualitative arguments, we performed a
quantitative MD simulation analysis. These Langevin based simulations have been used successfully
for studying isolated and correlated diffusion\cite{alexandrowicz_onset_2006,jardine_studying_2009}. Details of the simulation are given
in the supplementary information (SI) section. Figure 3a shows calculated $\alpha(\Delta K)$
curves for the two azimuths using a negligible
energy barrier for step crossing. The full green lines and dashed black lines mark the 0$^\circ$ and 67.5$^\circ$ azimuths respectively. As expected, the relatively isotropic
motion leads to similar $\alpha(\Delta K)$ curves for the two azimuths,
which are very different to the experimental results. Figure 3b shows calculations with a high energy
barrier which effectively blocks step crossing completely. The anisotropic
motion leads to a clear difference between the two azimuths,
supporting our qualitative arguments. However,
the initial correlation peak followed by the de Gennes narrowing dip,
related to the relative motion of Na atoms, does not follow the same simple scaling behavior seen in the experimental data
. Instead, along the 67.5 azimuth  we see a broad peak with a maximum
at approximately 1.2\AA$^{-1}$ (the vertical dashed lines mark
the positions of the two peaks in the experimental data, shown also in figures 2a and 2b).

An explanation for this discrepancy can be seen when comparing the
static structure factor (SSF), $I\left(\Delta\vec{k},t=0\right)$. The
SSF, which is simply the Fourier transform
of the average pair distribution function and effectively gives the same information as a diffraction pattern, is plotted on the right panels
of figure 3. For a negligible step-crossing barrier, the SSF (figure
3d), has a well defined peak at \textasciitilde{}0.8\AA$^{-1}$ 
reflecting a typical Na spacing of $2\pi /0.8=8$\AA. This peak appears roughly at the same position along the two
azimuths due to the almost isotropic motion obtained in the absence
of a significant step-crossing barrier. \footnote{A peak in the static
structure factor gives rise to a de Gennes narrowing dip in $\alpha(\Delta K)$
at roughly the same position; this relation can be derived from analytical
theory \cite{ala-nissila_collective_2002} and is also obvious when comparing figures
3a and 3d.} Figure 3e shows the SSF in the presence of
a significant step-crossing barrier. While there is a significant difference
between the two azimuths, the 67.5$^{\circ}$ azimuth still contains
a peak in the SSF at \textasciitilde{}0.8\AA$^{-1}$,
even though its intensity is smaller. The fact that our simulation
calculates a structure peak at approximately the same position along both azimuths refects the fact that we have used an isotropic Na-Na interaction
model which maintains the typical distances and
correlated motion between Na atoms on different terraces. 

Figures 3c and 3f show  $\alpha(\Delta K)$ and the
SSF when using a modified simulation which includes
a screening effect of the atomic steps, i.e. the repulsive force was set to zero between Na atoms  
on different terraces. As can be seen, making the interaction anisotropic
removes the static structure factor peak at \textasciitilde{}0.8\AA$^{-1}$
and results in a $\alpha(\Delta K)$ curve for the 67.5$^\circ$ azimuth which
looks like a simple stretch of the 0$^\circ$ azimuth, now consistent with the trend
in the experimental data.

A quantitative description of the diffusion
process was obtained by simultaneously fitting the measured
$\Delta K$, crystal azimuth, coverage and temperature dependencies (figures 2a,2b and 2c)
of the decay rates, $\alpha$.  The potential energy surface
we used allowed only one adsorption site within the width of the terrace
and has two adjustable
parameters, an energy barrier for moving along the steps, $E_{bridge}$,
and a barrier for step-crossing $E_{step}$. A repulsive dipolar interaction
model was used\cite{alexandrowicz_onset_2006} with the important modification, that
interactions were only allowed between Na atoms on the same terrace,
i.e. an anisotropic interaction.
The lines in figures 2a \& 2b show the simulated results for the best fit
parameters, $E_{bridge}=55meV$, $E_{step}\geq125meV$ ( higher
values did not make a noticeable difference). The simulated $\alpha$ values reproduce all the major
trends discussed above, and given the relatively simple model used and the minimal number of adjustable parameters the simulation mimics the data
surprisingly well. Given the good fit between the experiments and the MD simulation, we can use the later to calculate the hopping frequency of the Na atoms. These hopping rates, which are plotted in figure 2d, follow a simple Arrhenius behaviour with an activation energy of  $43 \pm 3$  meV and a pre-exponential factor of $2.7 \pm 0.6\cdot10^{12}$ sec$^{-1}$.  It is interesting to note that these hopping rates, which are predominantly in the, 0$^\circ$ azimuth, are significantly faster ($\approx$x4 at 155K) than on Na/Cu(001) at similar temperatures, reflecting the fact the barrier is lowered by 20 meV, resembling the macroscopic observations of CO on platinum surfaces.\cite{Ma1999L661}

In summary, from the lack of motion perpendicular to the surface plane and the comparison between the measured ISF along the two azimuths, we conclude that motion of the Na atoms involves only one adsorption site per terrace and that the motion is strongly oriented parallel to the step edges \footnote{It is important to note that we can not identify the exact position of the single adsorption site from which the diffusion takes place, i.e. whether it is located in the middle of the terrace or close to the upper or lower step, based on the diffusion data. Consequently, the choice of a PES with a minimum at the center of the terrace, as shown in the SI section, is based on simplicity and convenience.}. This observation could either indicate that there is only one minimum of the PES within the terrace width, or that Na atoms also reside in other adsorption sites which have significantly larger energy barriers (>120 meV) for diffusion and hence do not contribute to the measured dynamics. Future helium or electron diffraction studies, as well as measurements of the vibrational frequencies of Na/Cu(115) could potentially differentiate between these two scenarios.  Comparison with MD simulations reveals that not only do the atomic steps dramatically limit the diffusion across the steps (by
at least two orders of magnitude at 155K), and increase the rate of motion along the steps, they also effectively
shield the Na-Na interaction, making the spatial and temporal correlations
between Na atoms highly anisotropic as well. 

Atomic steps on bare metallic surfaces produce significant permanent dipoles\cite{park2005}. It seems plausible that Na atoms produce induced
dipoles at the step edges which oppose the repulsive force between Na atoms
on different terraces and lead to a loss of temporal and spatial
correlation. Since atomic steps are such a common surface defect and many atomic and molecular adsorbates experience dipolar interactions, our observation of how steps affect adsorbate diffusion is likely to be relevant to a wide range of surface systems. In particular, the screening effect of the atomic steps which makes the correlated motion anisotropic should be considered when designing vicinal surface templates for self organized nano-structures (e.g. \cite{cirera2014}).  Finally, Alkali atoms on copper surfaces also represent
a rather unique case where DFT calculations coincide with diffusion
data to a very high accuracy\cite{fratesi2009}. Hence, the atomic-scale diffusion measurements presented above provide a benchmark for testing how well DFT can explain and eventually predict the affect steps have on diffusion and on the interaction between adsorbates on surfaces. 

This work was supported by the German-Israeli Foundation for Scientific Research and Development, the Israeli Science Foundation
(Grant No. 2011185), the German Science Foundation (DFG) through contract MO 960/18-1, the Cluster of Excellence RESOLV (EXC 1069) and the European Research Council under the European
Union\textquoteright s seventh framework program (FP/2007-
2013)/ ERC Grant 307267.

Supporting Information Available ( http://pubs.acs.org/doi/suppl/10.1021/acs.jpclett.5b01939 )
(1) Description of the experimental methods (2) Angular scans along the $90^{\circ}$ azimuth, (3) An example of a measured ISF curve from which the decay rates are extracted (4) Description of the molecular dynamics simulation (5) MD simulations of a PES with two adsorption sites within the terrace width and (6) a movie illustrating the correlation of the trajectories under the non-isotropic interaction potential.


\pagebreak

\end{document}